\documentclass[12pt]{article}
\usepackage{graphicx}

\newcommand{\Frac}[2]{\frac{\displaystyle #1}{\displaystyle #2}}
\newcommand{\beq}{\begin{equation}}
\newcommand{\eeq}{\end{equation}}
\newcommand{\beqn}{\begin{eqnarray}}
\newcommand{\eeqn}{\end{eqnarray}}
\newcommand{\beqns}{\begin{eqnarray*}}
\newcommand{\eeqns}{\end{eqnarray*}}

\begin{document}
\begin{titlepage}
\begin{center}

\hfill USTC-ICTS-04-22\\

\vspace{2.5cm}

{\large {\bf Angular distribution asymmetry in
$\tau^-\to\pi^-\pi^0 \nu_\tau$ decay
in the two-Higgs-doublet model with large $\tan\beta$}}\\
\vspace*{1.0cm}
 {Dao-Neng Gao$^\dagger$} \vspace*{0.3cm} \\
{\it\small Interdisciplinary Center for Theoretical Study,
University of Science and Technology of China, Hefei, Anhui 230026
China
}

\vspace*{1cm}
\end{center}
\begin{abstract}
\noindent We study possible scalar type interactions in
$\tau^-\to\pi^-\pi^0\nu_\tau$ decay. One finds that an angular
distribution asymmetry, ${\cal A}(s)$, can be induced from the
interference between the scalar part and vector part amplitudes in
this decay. Our analysis shows that, in the two-Higgs-doublet
model of type II with large $\tan\beta$, the charged Higgs
contribution could make ${\cal A}(s)$  up to $4\times 10^{-3}$
without conflict with present experimental constraints. Thus in
the future precise experiments in $\tau$-charm factories, this
angular distribution asymmetry may be an interesting observable
either to be helpful in searching for the signal of the charged
Higgs boson or to impose the significant bound on it.
\end{abstract}

\vfill
\noindent

$^{\dagger}$ E-mail:~gaodn@ustc.edu.cn
\end{titlepage}

The $\tau$ lepton is the only known lepton massive enough to decay
into hadrons. Its semi-leptonic decay into a $\tau$ neutrino and a
hadron system provides an ideal tool for  testing the standard
model (SM), in both the electroweak and the strong sectors
\cite{Mar95, Pich97, Gan01}. With the increased experimental
sensitivities achieved already or in the future, some interesting
limits on possible new physics contributions to the $\tau$ decay
amplitudes may also be expected \cite{Pich97, Nelson96}. The main
purpose of the present paper is to explore this possibility in
$\tau^-\to\pi^-\pi^0\nu_\tau$ decay. In particular, within the
two-Higgs-doublet model (2HDM) we will analyze an angular
distribution asymmetry induced from this transition, and it is
interesting that this asymmetry may be very useful either to
search for the signal of the charged Higgs boson or to impose the
significant bound on it. Meanwhile, as a byproduct of our
calculation, we will show below that the problem with the pion
form factor \cite{AH04, Morse04} could not be solved by including
the charged Higgs contribution.

The final two-pseudoscalar mesons  in the decay
$\tau^-\to\pi^-\pi^0\nu_\tau$ could have spin-parity number
$J^{P}=0^{+}$ or $1^{-}$. Conservation of the vector current (CVC)
however forbids the production of $0^+$ non-strange states, thus
the decay is expected to be dominated by the low-lying vector
resonance contributions at low energies \cite{CLEO00, PP01}. It is
known that, in the limit of exact isospin symmetry, CVC relates
properties of the $\pi^-\pi^0$ system produced in $\tau$ decay to
those of the $\pi^+\pi^-$ produced in the reaction
$e^+e^-\to\pi^+\pi^-$, which implies the CVC relation
\beqn\label{CVC1}
\Frac{d\Gamma(\tau\to\pi^-\pi^0\nu_\tau)}{ds}=\Frac{3
\Gamma^{(0)}_e \cos^2\theta_C}{2\pi \alpha^2 m_\tau^2}s
(1-\frac{s}{m_\tau^2})^2(1+\frac{2s}{m_\tau^2})~\sigma_{e^+e^-\to\pi^+\pi^-}(s)\eeqn
with $$\Gamma^{(0)}_{e}=\frac{G_F^2 m_\tau^5}{192\pi^3},\;\;\;
s=(p_{\pi^-}+p_{\pi^0})^2, $$ $G_F$ is the Fermi coupling
constant, and $\theta_C$ is the Cabibbo angle. By comparison of
$e^+e^-\to\pi^+\pi^-$ data and the $\tau\to\pi^-\pi^0\nu_\tau$
data, one finds that the CVC relation (\ref{CVC1}) works very well
for the low $s$, except in the higher $s$ region \cite{CLEO00,
CVCdata, AH04}. It is thought that some of the discrepancy between
them may be understood by including the isospin symmetry breaking
effects \cite{DGJ}. However, it has been pointed out by A.
H\"ocker \cite{AH04}, corrections due to the isospin violation
from the SU(2)-breaking sources including the mass and width
differences of the charged and neutral $\rho(770)$ mesons, can
improve the agreement between $\tau$ and $e^+e^-$ data in the
$\rho$ peak region, while these cannot correct the discrepancy in
the tails, as shown in Fig. 1 of Ref. \cite{AH04}.

More recently, the author of Ref. \cite{Morse04} proposed that the
scalar contribution through the interference with the vector part
could be up to the percent level at $s\simeq$ 1 GeV$^2$,  which
may solve the above discrepancy. However, it will be shown below
this scalar type contribution has been overestimated in Ref.
\cite{Morse04}; but there is another interesting observable, the
angular distribution asymmetry, induced by the scalar contribution
in $\tau^-\to\pi^-\pi^0\nu_\tau$ decay. More interestingly, this
asymmetry may be enhanced in the 2HDM with large $\tan\beta$.

The general scalar and pseudoscalar type interactions in
$\tau^-\to\pi^-\pi^0\nu_\tau$ decay have been investigated by the
authors of Ref. \cite{TH93}, and the general invariant amplitude
for this decay, by assuming only left-handed neutrinos, can be
parameterized as \beqn\label{IA1}{\cal
M}=G_F\cos\theta_C~\left[F_V (p_{\pi^-}-p_{\pi^0})_\mu
\bar{u}(p_{\nu_\tau})\gamma^\mu(1-\gamma_5)u(p_\tau)\right.\nonumber
\\\left.+F_S m_\tau \bar{u}(p_{\nu_\tau})(1+\gamma_5)u(p_\tau)\right],\eeqn
where $F_V$ is the vector form factor, and $F_S$ the scalar one.
It is straightforward to get the differential decay rate
\beqn\label{width1} \frac{d\Gamma(\tau^-\to\pi^-\pi^0\nu_\tau)}{d
s}=\frac{\cos^2\theta_C\Gamma^{(0)}_e}{2m_\tau^2}\lambda^{1/2}(1,
m_{\pi^0}^2/s, m_{\pi^-}^2/s)
\left(1-\frac{s}{m_\tau^2}\right)\nonumber\\
\times \left\{|F_V|^2\left[\lambda(1,m_{\pi^0}^2/s, m_{\pi^-}^2/s
)\left(1+\frac{2s}{m_\tau^2}\right)+\frac{3(m_{\pi^-}^2-m_{\pi^0}^2)^2}{s^2}\right]\right.\nonumber\\
\left. +3|F_S|^2-6 {\rm Re}(F_V
F^*_S)\frac{m_{\pi^-}^2-m_{\pi^0}^2}{s}\right\},\eeqn where
$\lambda(a,b,c)=a^2+b^2+c^2-2 ab-2 bc-2 ca$. Since $F_S$ is very
small, in general one cannot expect the term proportional to
$|F_S|^2$ would give a significant contribution to $d\Gamma/d s$,
which should be below the percent level. It is easy to see that,
in the limit of the exact isospin symmetry $m_{\pi^-}=m_{\pi^0}$,
the interference term between the scalar and vector amplitudes in
eq. (\ref{width1}) will vanish. Using the experimental value,
$m_{\pi^-}-m_{\pi^0}=4.5936\pm 0.0005$ MeV \cite{pdg04}, we have
\beqn \frac{m_{\pi^-}^2-m_{\pi^0}^2}{s}\sim 10^{-3}\eeqn for
$s\simeq 1{\rm GeV}^2$. Therefore, contributions to $d\Gamma/d s$
from the scalar interaction cannot be expected to reach  the
percent level in $\tau^-\to\pi^-\pi^0\nu_\tau$ decay, which thus
disagrees with the conclusion obtained in Ref. \cite{Morse04} [our
following analysis in the 2HDM can explicitly lead to this
conclusion by using eqs. (\ref{width1}), (\ref{FV}), (\ref{FS}),
and the limits (\ref{bound})]. On the other hand, one can expect
another interesting observable induced from the interference
between the scalar and vector interactions in this decay. It is
seen that, in the limit of $m_{\pi^-}=m_{\pi^0}$, the differential
decay rate in terms of $s$ and $\theta$, the angle between the
three-momentum of $\pi^-$ and the three-momentum of $\tau^-$ in
the $\pi^-\pi^0$ rest frame, can be written as
\beqn\label{width2}\frac{d^2\Gamma}{d s ~d\cos\theta}=\frac{3
\Gamma^{(0)}_e \cos^2\theta_C}{4
m_\tau^2}\sqrt{1-\frac{4m_\pi^2}{s}}\left(1-\frac{s}{m_\tau^2}\right)^2\left\{|F_S|^2+|F_V|^2
\left(1-\frac{4m_\pi^2}{s}\right)
\right.\nonumber\\
\left.
\times\left[\frac{s}{m_\tau^2}+\left(1-\frac{s}{m_\tau^2}\right)\cos^2\theta\right]+2
{\rm Re}(F_V
F_S^*)\sqrt{1-\frac{4m_\pi^2}{s}}\cos\theta\right\},\eeqn and the
phase space is given by $$4 m_\pi^2\le s\le m_\tau^2,\;\;\;\;
-1\le\cos\theta\le 1 .$$ Note that the interference term between
$F_V$ and $F_S$ in eq. (\ref{width2}) is proportional to
$\cos\theta$, and will vanish after integrating over $\theta$ in
the full phase space, which is consistent with eq. (\ref{width1})
in the isospin limit. However, this term can lead to an angular
distribution asymmetry, which is defined as \beq\label{asym1}
 {\cal A}(s)=\Frac{\int^1_0 \left(\frac{d^2\Gamma}{dsd\cos\theta}\right)d\cos\theta-\int^0_{-1}
 \left(\frac{d^2\Gamma}{dsd\cos\theta}\right)d\cos\theta}{\int^1_0 \left(\frac{d^2\Gamma}
 {dsd\cos\theta}\right)d\cos\theta+\int^0_{-1}
 \left(\frac{d^2\Gamma}{dsd\cos\theta}\right)d\cos\theta}.\eeq
Thus together with eq. (\ref{width2}), we have \beq\label{asym2}
{\cal A
}(s)=\frac{3\Gamma^{(0)}_e\cos^2\theta_{C}}{2m_\tau^2}\left(1-\frac{4m_\pi^2}{s}\right)
\left(1-\frac{s}{m^2_\tau}\right)^2{\rm Re}(F_V
F_S^*)\left(\frac{d\Gamma}{d s}\right)^{-1},\eeq where
\beq\label{width3}\frac{d\Gamma}{d s}=\frac{\Gamma^{(0)}_e
\cos^2\theta_{C}}{2m_\tau^2}\sqrt{1-\frac{4m_\pi^2}{s}}\left(1-\frac{s}{m_\tau^2}\right)^2\nonumber\\
\left[|F_V|^2\left(1+\frac{2s}{m_\tau^2}\right)\left(1-\frac{4m_\pi^2}{s}\right)+3
|F_S|^2 \right].\eeq

It is known that, in the low energy region $\sqrt{s}\le 1$ GeV,
$F_V$ can be well described by the $\rho(770)$ meson dominance
\cite{CLEO00, PP01}, which reads
\beq\label{FV}F_V=\frac{m_\rho^2}{m_\rho^2-s-im_\rho\Gamma_\rho(s)}
\eeq with \beq \Gamma_\rho(s)=\frac{m_\rho s}{96\pi
f_\pi^2}\left\{\left({1-\frac{4m_\pi^2}{s}}\right)^{3/2}\theta(s-4m_\pi^2)+\frac{1}{2}\left(1-
\frac{4m_K^2}{s}\right)^{3/2}\theta(s-4m_K^2) \right\}.\eeq
 In order to get a significant asymmetry ${\cal A}(s)$ in
$\tau^-\to\pi^-\pi^0\nu_\tau$ decay, a sizable contribution to
$F_S$ must be generated. Theoretically, $F_S$ in the SM is an
isospin symmetry breaking effect \cite{CEN01}. Experimentally
there is no $(\pi\pi)$ scalar resonance observed so far in this
low energy region of this decay \cite{Morse04} (future precise
experiments are expected but not available yet so far). Large mass
scalar particles in the high energy region may give contributions
to $F_S$, however, which in general will be strongly suppressed by
the inverse of their large mass squared. Therefore it seems
difficult to observe the scalar effects in this decay in the low
energy region both from $d\Gamma/d s$ and from ${\cal A}(s)$. This
situation may be changed in the 2HDM however, in which a
significant charged Higgs contribution to $F_S$ could be expected
with the large value of $\tan\beta$.

In the minimal version of the SM only one Higgs doublet is
required, and a single physical neutral Higgs boson is left over
after spontaneous electroweak  symmetry breaking. The 2HDM is the
simplest extension of the SM with one extra Higgs doublet, which
contains three neutral and two charged Higgs bosons. To the
purpose of the present discussion, we shall work in the context of
the 2HDM of type II \cite{2HDM}, where two Higgs scalar doublets
($H_u$ and $H_d$) are coupled separately to the right-handed
up-type quarks, and the right-handed down-type quarks and the
charged leptons. On the other hand, the 2HDM of type II is
particularly interesting being the Higgs sector of the minimal
supersymmetric standard model (MSSM) \cite{MSSM}. In this case,
flavor-changing neutral current amplitudes are naturally absent at
the tree level \cite{GW77}, however, it is possible to accommodate
large down-type Yukawa couplings, provided that the ratio
$v_u/v_d=\tan\beta$, where $v_{u(d)}$ is the vacuum expectation
value of the Higgs doublet $H_{u(d)}$, is large.
Phenomenologically, there has been considerable interest in the
large $\tan\beta$ effects in $B$ decays such as $B\to\mu^+\mu^-$
\cite{BS}  due to the neutral Higgs contributions at the loop
level, as well as in the $\tau$ leptonic decays $\tau\to\ell
\nu_\ell \nu_\tau$ and semi-leptonic decays $\tau\to\pi/K
\nu_\tau$ due to the charged Higgs contributions starting from the
tree level \cite{tau}. The effects of the charged Higgs boson in
$\tau$ decays have also been studied in Ref. \cite{KP88}.

\begin{figure}[t]
\begin{center}
\includegraphics[width=7cm,height=4cm]{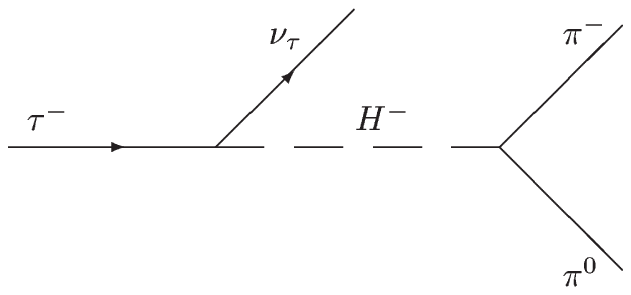}
\end{center}
\caption{Tree-level Feynman diagram for the charged Higgs
contribution to $\tau^-\to\pi^-\pi^0\nu_\tau$ decay.}
\end{figure}

The tree-level Yukawa interaction in the 2HDM of type II
(including the MSSM) can be written as \beq\label{yukuwa} {\cal
L}_Y=Y_d\bar{d}_RQ_L H_d+Y_u\bar{u}_RQ_L H_u+Y_L\bar{\ell}_R L
H_d+{\rm H.c.},\eeq where $Y_{u,d,\ell}$ are $3\times 3$ Yukawa
couplings matrices. Thus quarks and charged leptons together with
$W^\pm$ and $Z^0$ will get massive after spontaneous symmetry
breaking. Different from the SM case, now five physical Higgs
particles: two charged ones $H^\pm$ and three neutral ones $h^0$,
$H^0$, $A^0$, will be left over \cite{2HDM}. One can find that the
charged Higgs exchange will give the tree level contribution to
$\tau^-\to\pi^-\pi^0\nu_\tau$ decay (we do not think loop
contributions can significantly change our conclusion since we are
only interested in the order-of-magnitude estimate in the present
calculation), which is \beq\label{tree} {\cal
L}^{H^\pm}=\frac{G_F}{\sqrt{2}}\cos\theta_C
\frac{m_\tau\tan^2\beta}{m_{H^\pm}^2}\bar{\nu_\tau}(1+\gamma_5)\tau\left[m_d
\bar{d}(1-\gamma_5)u+\frac{m_u}{\tan^2\beta}\bar{d}(1+\gamma_5)u
\right],\eeq and the corresponding Feynman diagram has been drawn
in Fig. 1.  It is obvious that the scalar contribution to $F_S$
from the above ${\cal L}^{H^\pm}$ will be strongly suppressed by
$1/m^2_{H^\pm}$ for $m_{H^\pm}\sim O(\rm 10^2~GeV)$, which however
can be substantially compensated by large $\tan\beta$. In the
large $\tan\beta$ limit (so we can neglect the term proportional
to $m_u$), one has
\beq\label{FS}F_S=\frac{m_d\tan^2\beta}{m^2_{H^\pm}}\frac{m^2_\pi}{m_u+m_d}.\eeq
Unfortunately, at present there is no evidence for $m_{H^\pm}$
experimentally. From Ref. \cite{pdg04}, only the lower limit
$m_{H^\pm}> 79.3$ GeV is bounded, one can expect the possibility
of the significant $F_S$ for $\tan\beta\simeq 30\sim 50$. On the
other hand, some measurements have given the bounds on
$\tan\beta/m_{H^\pm}$ in the 2HDM of type II, which read
\beqn\label{bound} {m_{H^\pm}}>1.28~ \tan\beta~{\rm
GeV}~~~(95\%{\rm CL}),~\cite{OPAL03} \nonumber\\
{\tan\beta}/{m_{H^\pm}}<0.53~{\rm GeV^{-1}}~~~(95\%{\rm CL }),
~\cite{OPAL01}\\
{\tan\beta}/{m_{H^\pm}}<0.40~{\rm GeV^{-1}}~~~(90\%{\rm CL
}).~\cite{ALEPH01}\nonumber
 \eeqn
Using the above bounds and from eq. (\ref{FS}), we find that $F_S$
could be up to $10^{-3}$ in the 2HDM of type II with large
$\tan\beta$. Of course, this small value of $F_S$ could only give
the negligible contribution to $d\Gamma/d s$ defined in eq.
(\ref{width3}) or (\ref{width2}), however, it may lead to an
interesting angular distribution asymmetry ${\cal A}(s)$ in
$\tau^-\to\pi^-\pi^0\nu_\tau$ decay defined in eq. (\ref{asym2}).
To illustrate the order of the asymmetry ${\cal A}(s)$, we take
the most conservative bound listed in eq. (\ref{bound}),
$\tan\beta/m_{H^\pm}$=0.4 GeV$^{-1}$, and ${\cal A}(s)$ in the
range of 0.3 GeV$^2$$ \le s\le$ 1 GeV$^2$ has been plotted in Fig.
2 (we are interested in the low energy region, in which the
contribution to $F_V$ is almost saturated by the $\rho(770)$
meson). It is seen that this differential asymmetry could be up to
$4\times 10^{-3}$ in the 2HDM of type II at large $\tan\beta$
without conflict with the present experimental constraints, which
may be detected in the future precise experiments in $\tau$-charm
factories. Note that ${\cal A}(s)$ defined in eq. (\ref{asym2}) is
proportional to ${\rm Re}(F_V F_S^*)$, and from eq. (\ref{FV}),
\beq {\rm
Re}(F_V)=\frac{m_\rho^2(m_\rho^2-s)}{(m_\rho^2-s)^2+m_\rho^2\Gamma^2_\rho(s)},\eeq
which vanishes for $s=m_\rho^2$, thus the sign of the differential
asymmetry will be changed (as shown in Fig. 2), and the integrated
asymmetry over $s$ is not very significant.

\begin{figure}[t]
\begin{center}
\includegraphics[width=12cm,height=9cm]{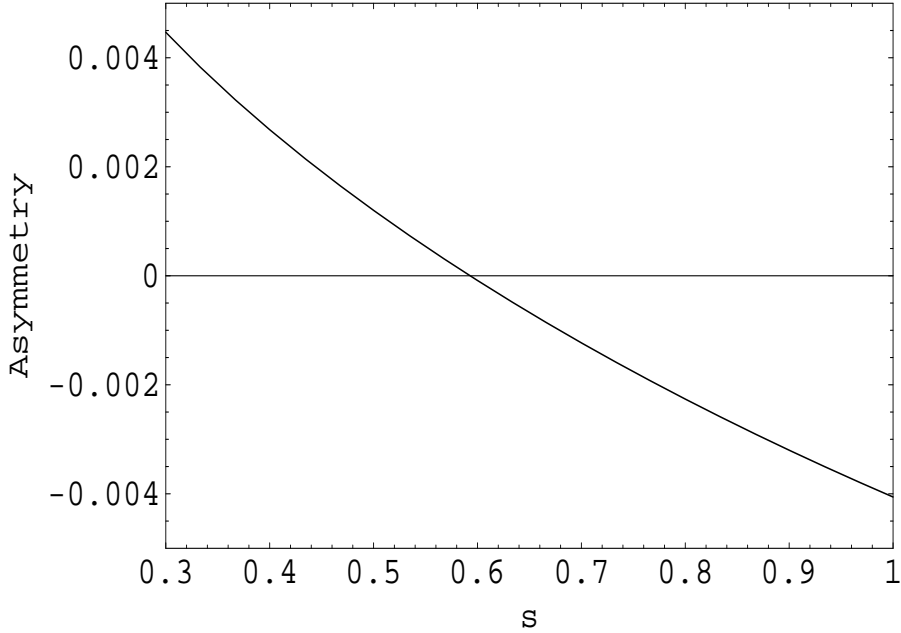}
\end{center}
\caption{ The differential angular distribution asymmetry ${\cal
A}(s)$ for 0.3 GeV$^2$$ \le s\le$ 1 GeV$^2$ with
$\tan\beta/m_{H^\pm}=0.4$ GeV$^{-1}$.}
\end{figure}

We have analyzed the decay of $\tau^-\to\pi^-\pi^0\nu_\tau$  by
considering possible scalar type interactions. We find that the
inclusion of scalar type interactions cannot still explain the
present discrepancy from the data between
$\sigma(e^+e^-\to\pi^+\pi^-)$ and
$d\Gamma(\tau^-\to\pi^-\pi^0\nu_\tau)/ds$ if it really exists,
which disagrees with the conclusion obtained in Ref.
\cite{Morse04}. However, scalar type interactions can lead to an
angular distribution asymmetry in $\tau^-\to\pi^-\pi^0\nu_\tau$
decay. Interestingly, due to the charged Higgs contribution in the
2HDM of type II with large $\tan\beta$, present experimental
constraints allows that the differential asymmetry ${\cal A}(s)$
could be up to $4\times 10^{-3}$, thus it is expected that the
future precise measurements of this asymmetry may either help to
search for the signal of the charged Higgs boson or impose the
significant bound on it.

\vspace{0.5cm}
\section*{Acknowledgements}
The author wishes to thank W.M. Morse for very helpful
communications. This work was supported in part by the National
Natural Science Foundation of China under Grant No. 10275059 and
supported in part by the Alexander von Humboldt Foundation.

\end{document}